\documentclass[a4paper]{jpconf}
\usepackage{graphicx}

\usepackage{citesort}

\begin{document}
\title{Explosive nucleosynthesis in core-collapse supernovae}

\author{Almudena Arcones}

\address{Department of Physics, University of Basel,
    Klingelbergstra{\ss}e 82, 4056, Basel, Switzerland}

\ead{a.arcones@unibas.ch}

\begin{abstract}
  The specific mechanism and astrophysical site for the production of
  half of the elements heavier than iron via rapid neutron capture
  (r-process) remains to be found.  In order to reproduce the
  abundances of the solar system and of the old halo stars, at least
  two components are required: the heavy r-process nuclei ($A>130$)
  and the weak r-process which correspond to the lighter heavy nuclei
  ($A<130$). In this work, we present nucleosynthesis studies based on
  trajectories of hydrodynamical simulations for core-collapse
  supernovae and their subsequent neutrino-driven winds.  We show that
  the weak r-process elements can be produced in neutrino-driven winds
  and we relate their abundances to the neutrino emission from the
  nascent neutron star.  Based on the latest hydrodynamical
  simulations, heavy r-process elements cannot be synthesized in the
  neutrino-driven winds. However, by artificially increasing the wind
  entropy, elements up to $A=195$ can be made. In this way one can
  mimic the general behavior of an ejecta where the r-process
  occurs. We use this to study the impact of the nuclear physics input
  (nuclear masses, neutron capture cross sections, and beta-delayed
  neutron emission) and of the long-time dynamical evolution on the
  final abundances.
\end{abstract}

\section{Introduction}
Half of the elements heavier than iron are produced by rapid neutron
captures in a yet unknown astrophysical scenario. After the initial
success of \cite{Woosley94} in reproducing observed solar r-process
abundances, core-collapse supernovae and the subsequent
neutrino-driven winds became one of the most promising candidates for
the production of r-process elements because their extreme explosive
conditions are very close to the ones needed for the r-process (see
e.g.,~\cite{Hoffman97,Thompson.Burrows.Meyer:2001,Otsuki.Tagoshi.ea:2000}).
Moreover, galactic chemical evolution models favor core-collapse
supernovae, since they occur early and frequently enough to account
for the abundances observed in old halo stars and in the solar system
\cite{Ishimaru.Wanajo:1999,Ishimaru.etal:2004}.  Although the
necessary conditions to produce heavy elements ($A>130$) are
identified \cite{Meyer92} (high entropies, low electron fractions, and
short expansion timescales), these are not found in the most recent
long-time supernova simulations \cite{Pruet.Hoffman.ea:2006,%
  arcones.janka.scheck:2007,Wanajo.Nomoto.ea:2009,%
  Fischer.etal:2010,Huedepohl.etal:2010}.

When a supernova explodes, matter surrounding the proto-neutron star
is heated by neutrinos and expands very fast reaching sometimes even
supersonic velocity
\cite{duncan.shapiro.wasserman:1986,Thompson.Burrows.Meyer:2001}.
This neutrino-driven wind moves through the early supernova ejecta and
eventually collides with it. The interaction of the wind with the
slow-moving ejecta results in a wind termination shock or reverse
shock where kinetic energy is transformed into internal
energy. Therefore, the expansion velocity drops and the temperature
(and thus the entropy) increases after the reverse shock.  The matter
near the proto-neutron star consists mainly of neutrons and protons
due to the high temperatures in this region.  When a mass element
expands, its temperature decreases and neutrons and protons recombine
to form alpha particles. The density also decreases but the
triple-alpha reaction combined with different alpha capture reactions
are still operating, resulting in heavy seed nuclei
\cite{Woosley.Hoffman:1992,Witti.Janka.Takahashi:1994}. The evolution
once the alpha particles start forming heavier nuclei depends on the
neutron-to-seed ratio.

The results presented here are based on our hydrodynamic simulations
\cite{arcones.janka.scheck:2007} where the neutron-to-seed ratio is
too low for the r-process to produce elements up to the third peak
($A=195$). In Sect.~\ref{sec:lepp} the nucleosynthesis obtained from
such simulations is discussed.  In Sect.~\ref{sec:rprocess} we have
used the neutrino-driven wind simulations with the entropy
artificially increased to study the impact of the long-time evolution
and nuclear physics input on the dynamical r-process. More details can
be found in \cite{Arcones.Montes:2010,Arcones.Martinez-Pinedo:2010}.

\section{Supernova simulations and nucleosynthesis networks}
\label{sec:sn_netw}
The investigatigation of the nucleosynthesis in neutrino-driven winds
is done in two steps. First, the evolution of the supernova ejecta is
followed during several seconds with hydrodynamical
simulations. Second, the composition is calculated by means of
extended nuclear reaction networks which include nuclei from stability
to both drip lines, reaching thus regions where no experimental data
are available and theoretical predictions are quite uncertain.

The modeling of the supernova ejecta during several seconds after
explosion is currently difficult since the supernova explosion
mechanism is not yet well understood \cite{Janka.Langanke.ea:2007} and
it is computationally expensive to perform long-time,
multidimensional, systematic studies for different progenitor stars,
as would be desirable in nucleosynthesis studies. Possible ways to
overcome this limitation include using parametric steady-state wind
models (e.g.,~\cite{Thompson.Burrows.Meyer:2001}) and forcing an
explosion by artificially changing neutrino properties
\cite{arcones.janka.scheck:2007,Fischer.etal:2010}. The evolution of
the outflow is rather independent of the details of the explosion
mechanism, but depends more on the evolution of the neutron star and
on the neutrino emission. Therefore, such approximations are a good
basis for nucleosynthesis studies. Although steady-state wind models
cannot consistently describe hydrodynamical effects (e.g., reverse
shock or multidimensional instabilities), both approaches agree in the
wind phase \cite{Arcones.Janka:2010}.

For our nucleosynthesis studies we use trajectories, i.e. density and
temperature evolutions, from
Ref.~\cite{arcones.janka.scheck:2007}. The composition is calculated
initially by assuming nuclear statistical equilibrium at $T=10$~GK,
almost only nucleons and few alpha particles are present. The
evolution of the composition is then followed using a full reaction
network \cite{Froehlich.Martinez-Pinedo.ea:2006}, which includes
nuclei from H to Hf with both neutron- and proton-rich isotopes.
Reactions with neutral and charged particles were taken from the
calculations of the statistical code NON-SMOKER
\cite{Rauscher.Thielemann:2000} and experimental rates were included
(NACRE, \cite{Angulo.Arnould.ea:1999}) when available. The theoretical
weak interaction rates are the same as in
Ref.~\cite{Froehlich.Martinez-Pinedo.ea:2006}.  When the conditions of
the supernova outflow are favourable for the r-process, i.e. high
neutron-to-seed ratio, we use a r-process network after
charged-particle reactions freeze out.  This network, which includes
photodissotiation, neutron capture, beta decay, and fission, is fully
implicit. Therefore, it can be used to study the late evolution when
matter decays to stability and the neutron density becomes very low.

\section{The origin of LEPP nuclei in supernovae}
\label{sec:lepp}
Most of the recent progress in understanding the origin of elements
commonly associated with the r-process is due to observations of ultra
metal-poor (UMP) stars (see \cite{Sneden.etal:2008} for recent
review).  The elemental abundances observed in the atmosphere of these
very old stars come from a few events.  These stars generally present
a robust pattern for ``heavy'' elements $56<Z<83$, in agreement with
the expected contribution of the r-process to the solar system, but
show some scatter for ``light'' elements $Z<47$
\cite{Sneden.etal:2008}. This suggests that at least two types of
events contribute to the r-process abundances
\cite{Wasserburg.etal:1996,Qian.Wasserburg:2001,Travaglio.Gallino.ea:2004,Aoki.etal:2005,%
  Otsuki.etal:2006,Montes.etal:2007}.  Qian and Wasserburg
\cite{Qian.Wasserburg:2007} argued that supernovae from low-mass
progenitors with $8 M_\odot< M < 12 M_\odot$ lead to all ``heavy'' and
some ``light'' elements, and that explosions of more massive
progenitors, $12 M_\odot< M < 25 M_\odot$, contribute to the remaining
light $A<130$ elements.

The process leading to elements with $A<130$ has been called in the
literature the weak r-process \cite{Wasserburg.Busso.Gallino:1996},
charged-particle reaction (CPR) process \cite{Woosley.Hoffman:1992,%
  Freiburghaus.Rembges.ea:1999,Qian.Wasserburg:2007}, and Light
Elemental Primary Process (LEPP)
\cite{Travaglio.Gallino.ea:2004,Montes.etal:2007}. We refer to this as
LEPP because such name does not make any reference to the specific
nuclear reactions or astrophysical environment. The term LEPP was
first introduced in Ref.~\cite{Travaglio.Gallino.ea:2004} which used a
galactic chemical evolution model to search for possible astrophysical
environments producing the elements such as Sr, Y, and Zr. Taking into
account the standard s-process and r-process contributions, they found
that non-negligible abundances of several isotopes ($^{86}$Sr,
$^{93}$Nb, $^{96}$Mo, $^{100}$Ru, $^{104}$Pd, $^{110}$Cd) were still
unexplained. They argued that a new ``light element primary process''
may have produced them and they discussed where and how this LEPP
could occur in order to explain their solar system abundances.  Montes
et al. in Ref.~\cite{Montes.etal:2007} suggested that the LEPP
observed in UMP stars show a hint of robustness of the process and
could have contributed to the solar system abundances.

\begin{figure}[!htb]
  \begin{tabular}{cc}
    \includegraphics[width=0.47\linewidth]{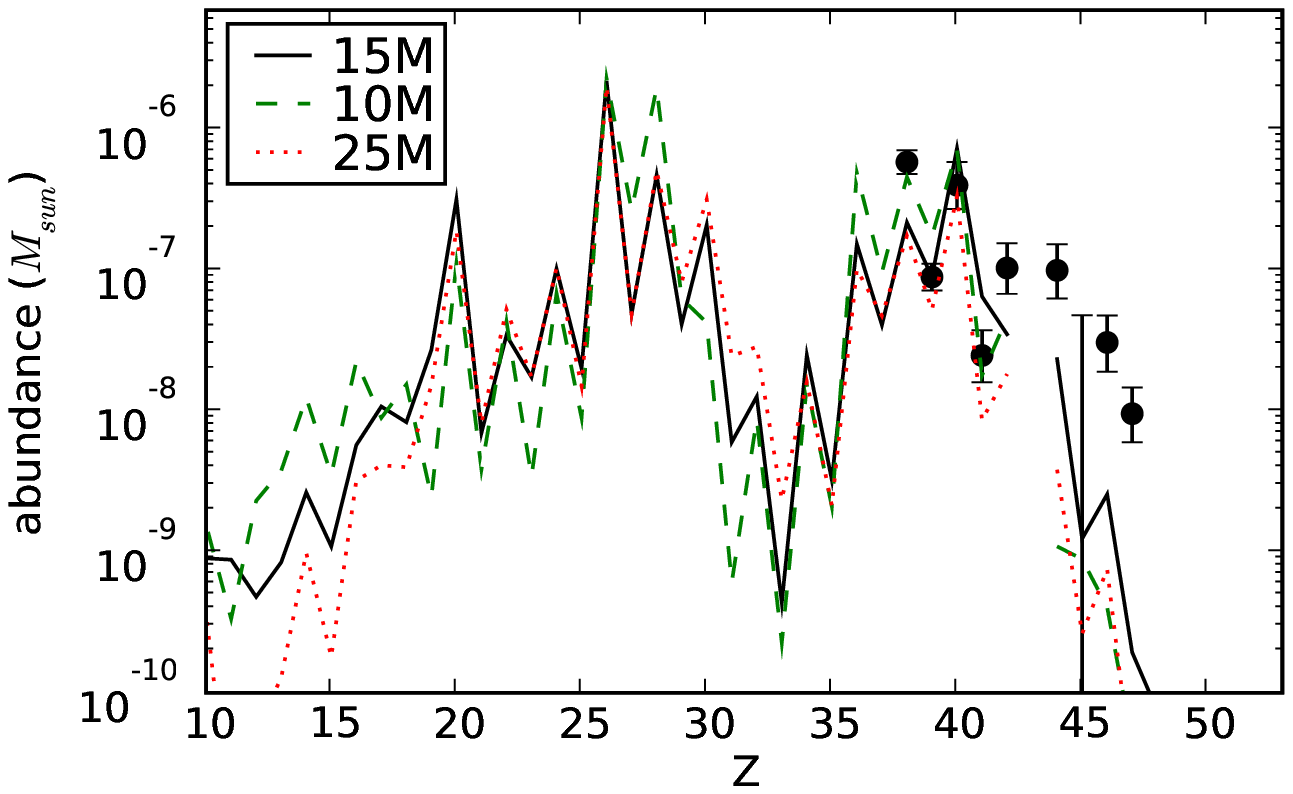}&
    \includegraphics[width=0.47\linewidth]{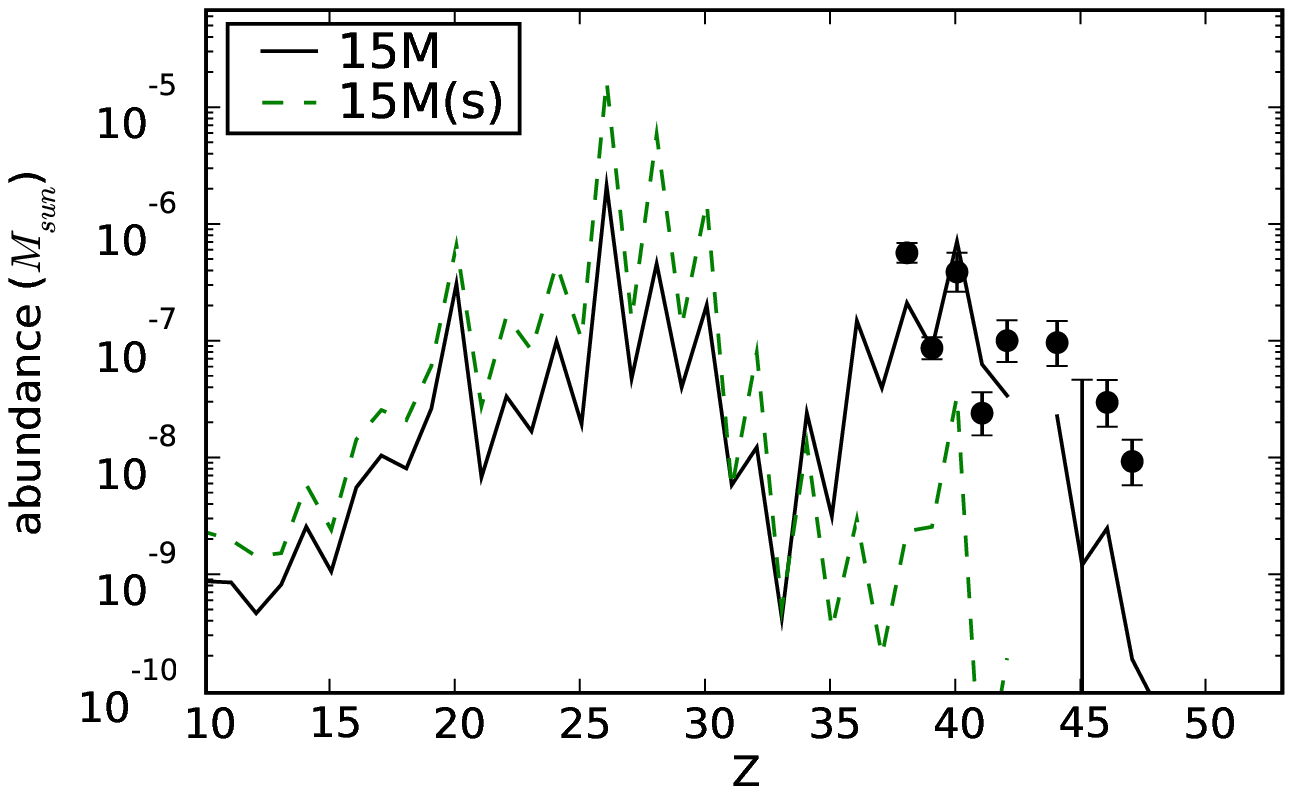}
  \end{tabular}
  \caption{Integrated abundances compared to the LEPP pattern
    \cite{Qian.Wasserburg:2008} rescaled to Z=39. The abundances of
    different progenitors with a similar evolution of the
    proto-neutron star are shown in the left panel, while the right
    panel gives the abundances of the same progenitor with different
    proto-neutron star evolution.}
  \label{fig:yesim}
\end{figure}

\begin{figure}[!htb]
  \includegraphics[width=0.45\linewidth]{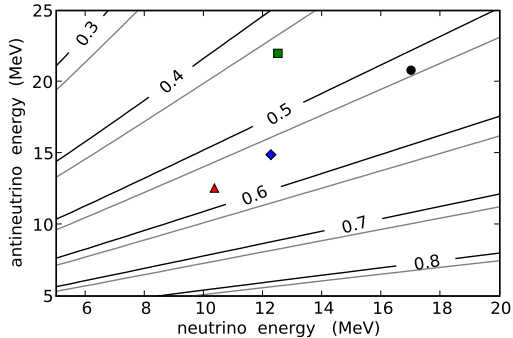}\hspace{1pc}%
  \begin{minipage}[b]{0.5\linewidth}\caption{ \label{fig:neut2col}
      Contours represent the electron fraction based on the
      approximation of \cite{Qian.Woosley:1996} (black contours:
      $L_{\bar{\nu}_e}/L_{\nu_e}=1$, grey contours:
      $L_{\bar{\nu}_e}/L_{\nu_e}=1.1$). The points indicate
      approximately the electron neutrino and antineutrino energies
      for different supernova models: the green square from
      \cite{Woosley.Wilson.ea:1994}, the black circle from model
      M15-l1-r6 of \cite{arcones.janka.scheck:2007}, the red triangle
      is from a 10~$M_{\odot}$ progenitor of \cite{Fischer.etal:2010},
      and the blue diamond from \cite{Huedepohl.etal:2010}, all at 10~s
      after bounce.\vspace{1pc}}
\end{minipage}
\end{figure}

\begin{figure}[!htb]
    \includegraphics[width=0.6\linewidth]{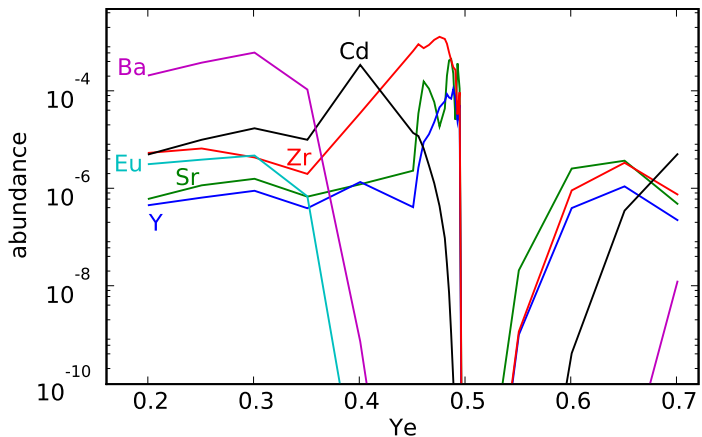}\hspace{1pc}%
    \begin{minipage}[b]{0.3\linewidth}\caption{\label{fig:abund_ye}
        Dependence of the abundances of representative elements (Sr,
        Y, Zr, Cd, Ba and Eu) on the electron fraction. These
        abundances result from a mass element ejected at 5s after the
        explosion in model 15M.\vspace{2pc}}
\end{minipage}  
\end{figure}

Figure~\ref{fig:yesim} shows the integrated nucleosynthesis for
different stellar progenitors with masses of 10,~15,~and~25~$M_\odot$,
as well as for a model with low wind entropy and slow expansion (due
to the less compact neutron star) which is labeled as 15M(s). Our
results confirm that no heavy r-process elements can be synthesized in
such explosions, however the LEPP can be realized based on the
simulations and for a range of realistic conditions. By comparing
models 15M and 15M(s), we explore the known dependence of the
nucleosynthesis on the entropy and expansion timescale
\cite{Qian.Woosley:1996}. The third wind parameter is the electron
fraction which is given by the neutrino properties determined by
neutrino interactions and transport. Therefore, the exact calculation
of the electron fraction remains a very challenging open problem
\cite{Huedepohl.etal:2010}. Figure~\ref{fig:neut2col} shows the
electron neutrino and antineutrino energies for different supernova
models. The antineutrino energy has decrease as the neutrino reactions
and transport have been improved leading to proton-rich winds in the
most recent simulations as shown by the $Y_e$ contours. This motivated
our exploration of the impact of the electron fraction on the
production of the LEPP elements. Figure~\ref{fig:abund_ye} illustrates
that the LEPP elements can be obtained for different proton- and
neutron-rich conditions.

\begin{figure}[!htb]
  \includegraphics[width=0.47\linewidth]{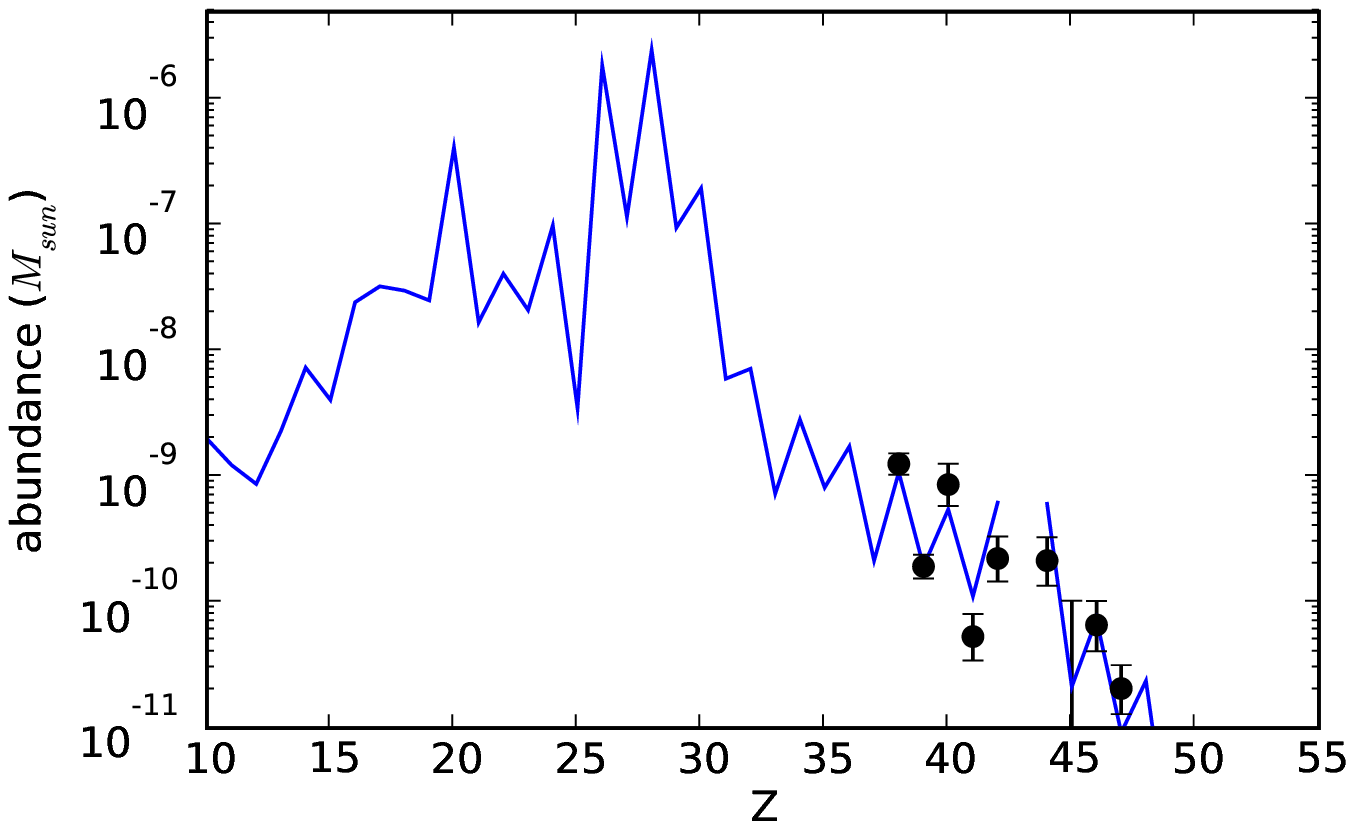}%
  \includegraphics[width=0.47\linewidth]{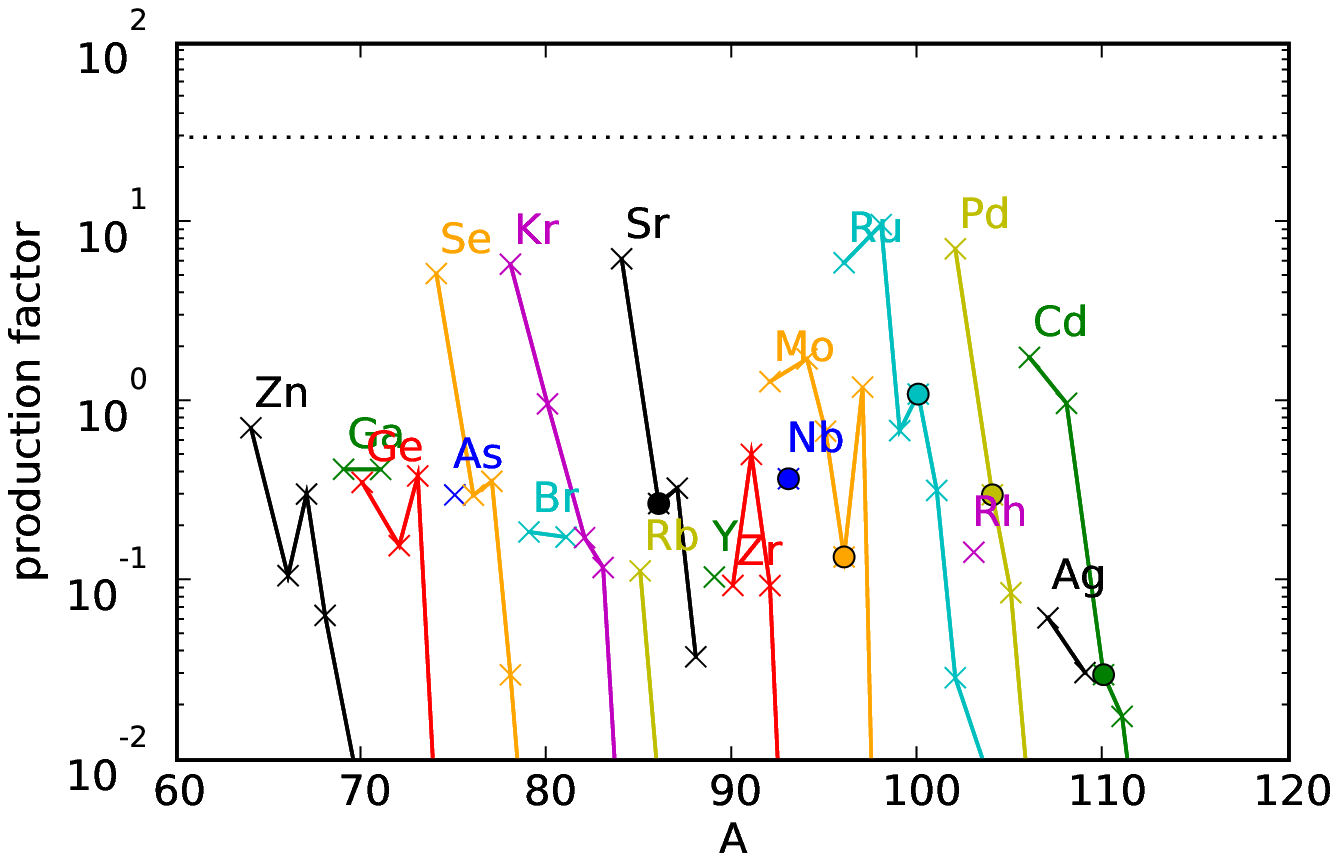}
  \caption{Integrated abundances (left panel) for mass elements
    ejected during the first ten seconds after the explosion of model
    15M. The electron fraction evolution is taken to evolve from
    $Y_e=0.5$ at 1s to $Y_e=0.65$ at 10s following
    \cite{Huedepohl.etal:2010}.  The right panel shows the
    corresponding production factors which are well below the
    overproduction limit marked by the dotted line and that almost
    only p-nuclei are synthesized. The isotopes marked with circles
    are not produced in enough quantities in the model of
    Ref.~\cite{Travaglio.Gallino.ea:2004} and are expected to be
    produced by the LEPP.}
  \label{fig:prich}
\end{figure}

Left panel in Fig.~\ref{fig:prich} shows that the LEPP pattern (dots
obtained from observation as in Ref.~\cite{Qian.Wasserburg:2008}) is
reproduced in proton-rich winds. Moreover, we found that this
abundance pattern is quite robust under small variations of the
evolution of $Y_e$ and of the wind parameters (e.g. changes of 20\% in
the entropy). However, elements heavier than iron-group nuclei can be
produced only when the neutrino fluxes are high enough to allow a
successful $\nu$p-process
\cite{Froehlich.Martinez-Pinedo.ea:2006,Pruet.Hoffman.ea:2006,Wanajo:2006}.
The right panel of Fig.~\ref{fig:prich} shows the production factor
for various isotopes (see \cite{Arcones.Montes:2010} for detailed
discussion). Although there is no overproduction (i.e. production
factors are below dotted line in Fig.~\ref{fig:prich}) and the
elemental abundances nicely reproduce the observed LEPP pattern in UMP
stars, almost only neutron-deficient isotopes (p-nuclei) are
produced. Therefore, proton-rich conditions can explain the LEPP
elements observed in UMP stars but not the missing isotopic abundances
in the solar system \cite{Travaglio.Gallino.ea:2004}. An exciting
possibility of proton-rich winds is the synthesis of the light
p-nuclei, since mainly neutron-deficient isotopes are present in the
wind ejecta
\cite{Froehlich.Hauser.ea:2006,Froehlich.Martinez-Pinedo.ea:2006,Wanajo.etal:2010}.

When the electron fraction is assumed to evolve towards neutron-rich
conditions, the LEPP pattern can be also reproduced but it is very
sensible under variations of $Y_e$ or of the wind parameters.  This
scenario may contribute to the LEPP elements found in the solar system
abundances because neutron-rich isotopes are produced. However, we
find an overproduction around $A \sim 90$ that was already pointed out
in previous nucleosynthesis studies based on supernova simulations
(see e.g.,\cite{Hoffman.Woosley.ea:1996}). This overproduction problem
and the fact that most recent supernova simulations
\cite{Fischer.etal:2010,Huedepohl.etal:2010} favor proton-rich winds
could suggest that neutron-rich winds are rare events.

Observation of isotopic abundances in UMP stars are very promising to
constraint the neutron richness of the neutrino-driven wind and thus
the evolution of the electron fraction and the neutrino properties in
supernovae.

\section{Impact of the nuclear physics input on the dynamical
  r-process}
\label{sec:rprocess}
We investigate the sensitivity of r-process abundances to the combined
effects of the long-time dynamical evolution and nuclear physics input
and provide a link between the behaviour of nuclear masses far from
stability and features in the final abundances.  The trajectory for
this study is also from the neutrino-driven wind simulations of
Ref.~\cite{arcones.janka.scheck:2007} where no heavy r-process
elements can be synthesized \cite{Arcones.Montes:2010}. Therefore, we
need to artificially increase the neutron-to-seed ratio (by increasing
the entropy by a factor two) in order to produce the third r-process
peak. This allows us to study the nucleosynthesis of heavy elements in
a typical high-entropy neutrino-driven wind in a more consistent way
than with fully parametric expansions
\cite{Freiburghaus.Rembges.ea:1999,Farouqi.etal:2010} or with
steady-state wind models (see e.g.,
\cite{Otsuki.Tagoshi.ea:2000,Thompson.Burrows.Meyer:2001}), which
cannot consistently explore the interaction of the wind with the slow
supernova ejecta that results in a reverse shock.

\begin{figure}[!htp]
  \includegraphics[width=0.47\linewidth]{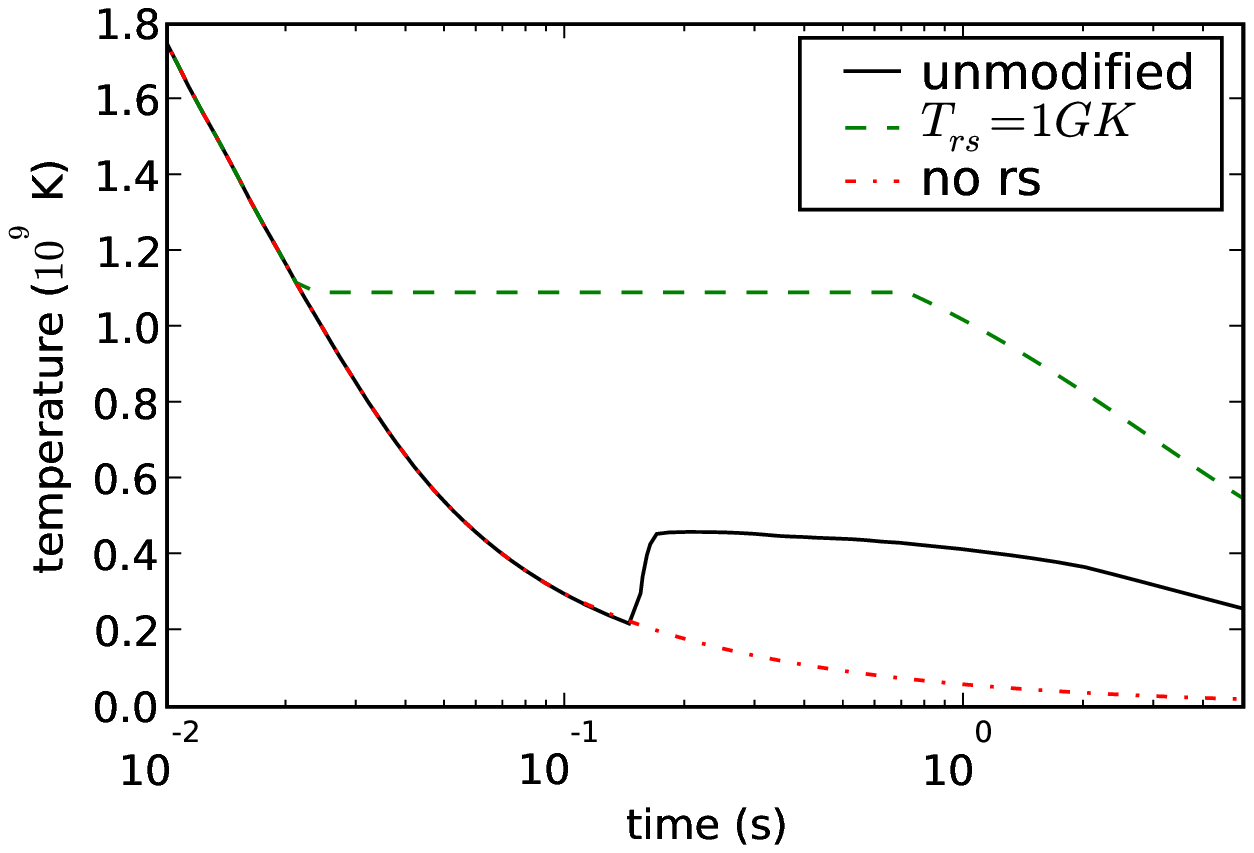}%
  \includegraphics[width=0.47\linewidth]{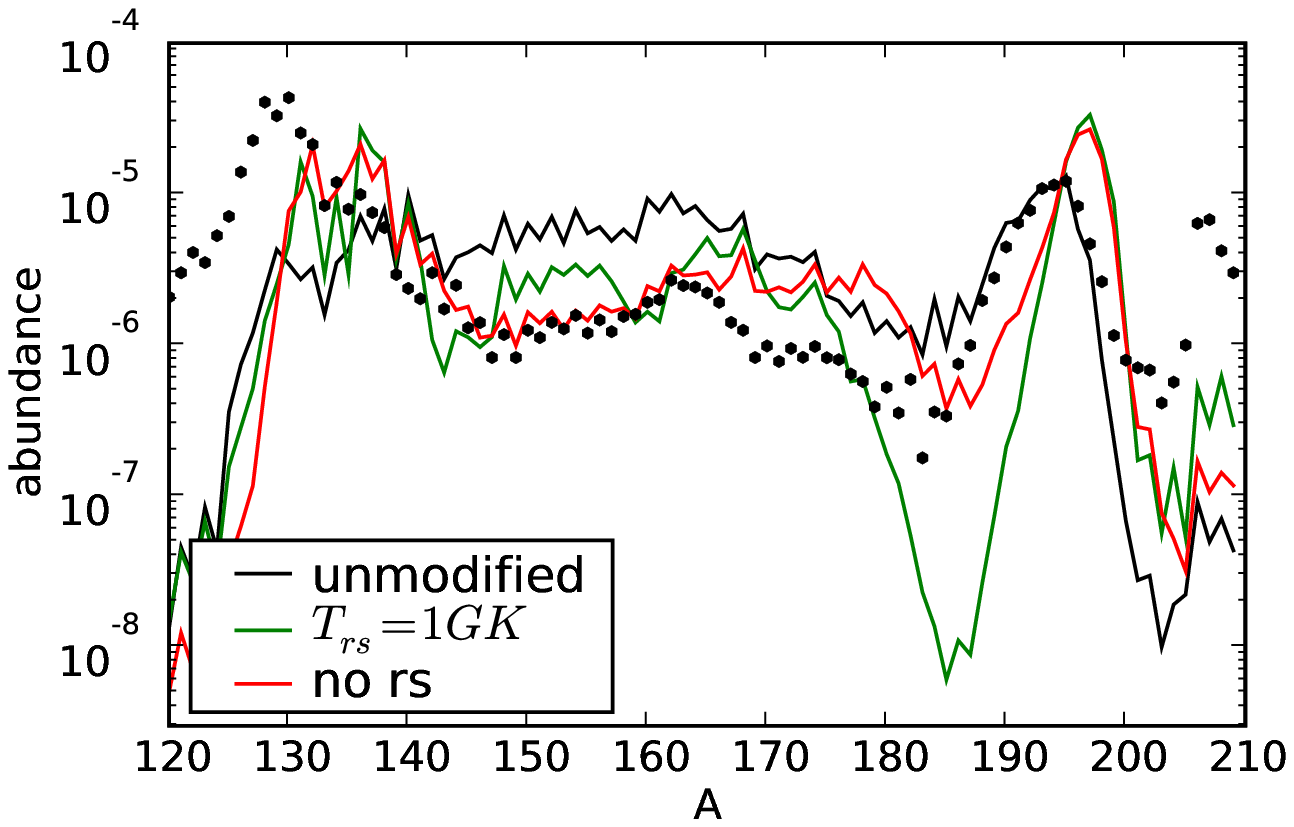}
  \caption{Temperature evolution (left panel) of a mass element
    ejected at 8~s after the explosion and variations of the long-time
    evolution. The right panel shows the final abundances (based on
    ETFSI-Q mass model) for the three trajectories and the solar
    abundances by dots.}
  \label{fig:rs_td}
\end{figure}

The evolution of temperature and density during the alpha-process
determines the neutron-to-seed ratio and thus the possibility of
forming heavy elements. However, the dynamical evolution after the
freeze-out of charged-particle reactions is affecting the final
abundances. In the left panel of Fig.~\ref{fig:rs_td} we present the
three trajectories used for our calculations. The trajectory labeled
as ``unmodified'' correspond to the hydrodynamical simulations with
the entropy increased and the reverse shock as in the simulations. We
change the position of the reverse shock to investigate the effect of
the variation of the dynamical evolution during the r-process, keeping
the same initial neutron-to-seed. In the trajectory labeled as
$T_{\mathrm{rs}}=1$~GK the reverse shock is at high temperatures,
while in the one labeled as ``no rs'' there is no reverse shock. The
abundances resulting from these three evolution are shown in the right
panel of Fig.~\ref{fig:rs_td}, compared to the solar abundances shown
by dots. Notice that the long time evolution has a big impact on the
position of the peaks and on the troughs.

\begin{figure}[!htp]
  \includegraphics[width=0.47\linewidth]{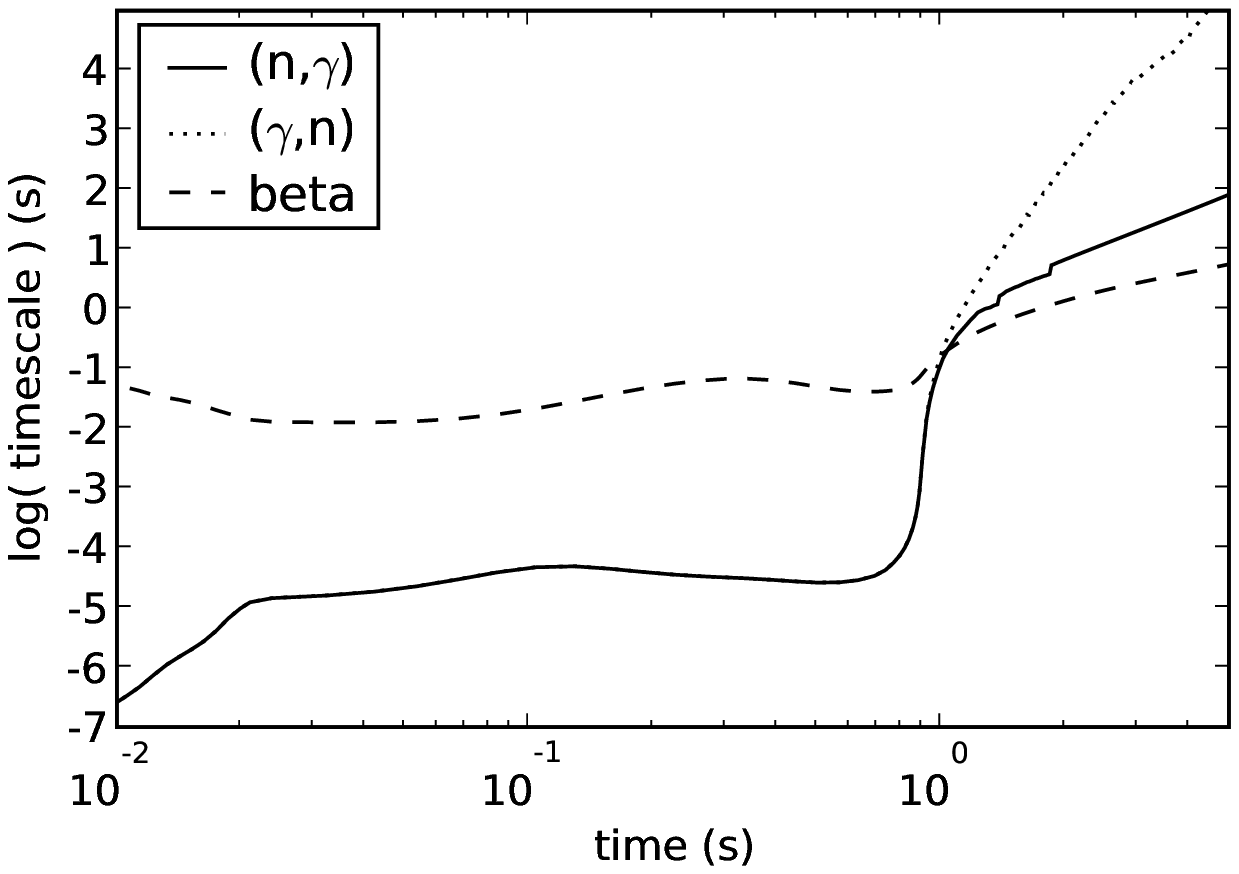}%
  \includegraphics[width=0.47\linewidth]{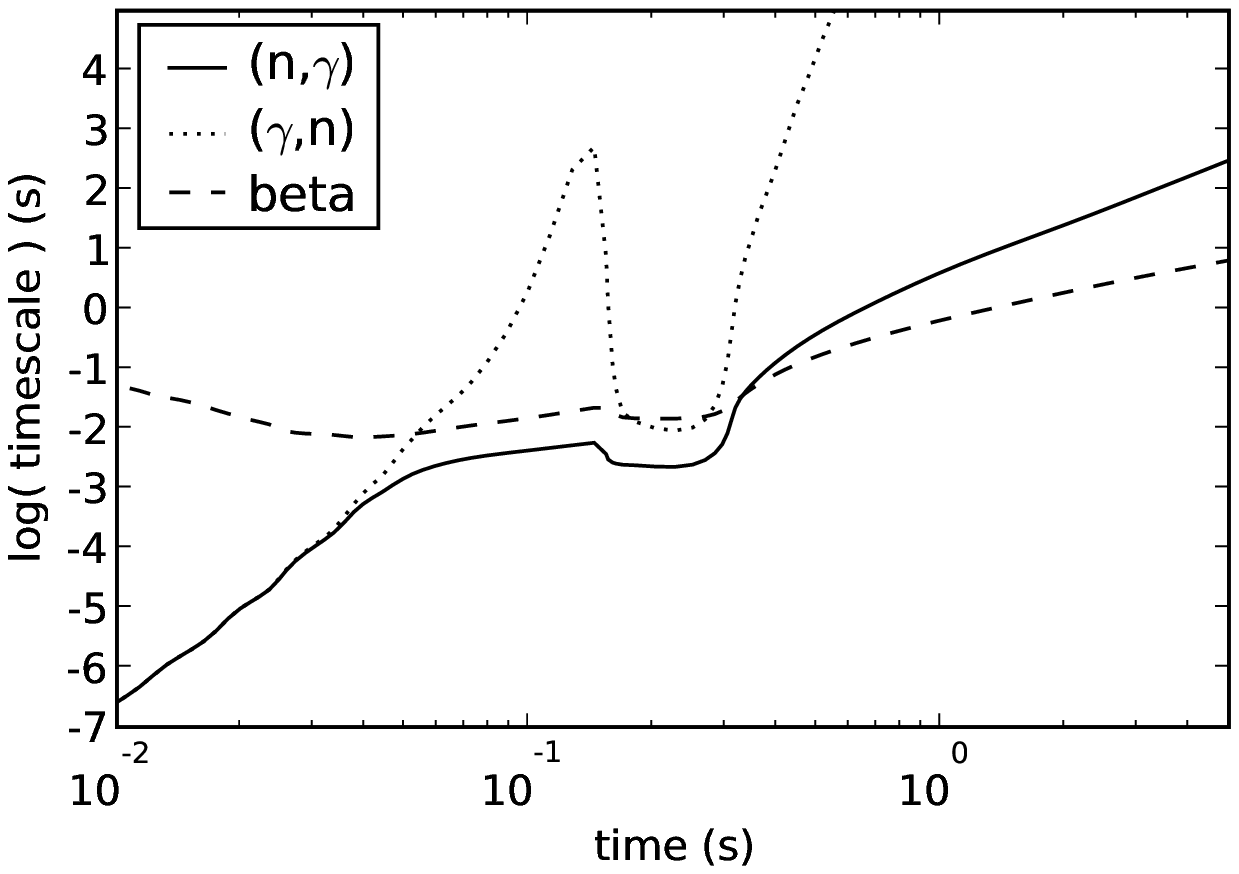}
  \caption{Evolution of relevant timescales for the trajectories in
    Fig.~\ref{fig:rs_td} labeled as ``$T_{\mathrm{rs}}=1$GK''(left
    panel) and ``unmodified'' (right panel). Solid, dashed, and dotted
    lines represent neutron capture, beta decay, and photodissociation
    timescales, respectively.}
  \label{fig:rs_tau}
\end{figure}

When the reverse shock is at high temperatures the evolution proceeds
under a $(n,\gamma )$--$ ( \gamma,n)$ equilibrium which lasts until
neutrons are exhausted. This equilibrium evolution is similar to the
classical r-process \cite{Kratz.Bitouzet.ea:1993} and it is also known
as hot r-process \cite{Wanajo:2007}. If the evolution proceeds at low
temperatures ($T<0.5$GK), there is a competition between neutron
capture and beta decay. This non-equilibrium evolution correspond to
the cold r-process introduced in
Ref.~\cite{Wanajo:2007}. Figure~\ref{fig:rs_tau} shows the evolution
of timescales for the three relevant processes: neutron capture,
photodissociation, and beta decay. The main difference between hot
(left panel) and cold (right panel) r-process is that in the latter
photodissociation is negligible. Therefore, the r-process path can
move farther away from stability reaching nuclei with shorter
half-lives and leading to a faster evolution and an earlier freeze
out. Moreover, neutron separation energies have less impact on the
final abundances because they enter only through the neutron capture
cross section. Notice that photodissociation depends exponentially on
the neutron separation energy. The importance of the different nuclear
physics input depends thus on the dynamical evolution, therefore all
our studies are performed with the equilibrium and non-equilibrium
evolutions \cite{Arcones.Martinez-Pinedo:2010}.

\subsection{Sensitivity to the mass model}
\label{sec:mm}
\begin{figure}[!htp]
  \includegraphics[width=0.51\linewidth,angle=0]{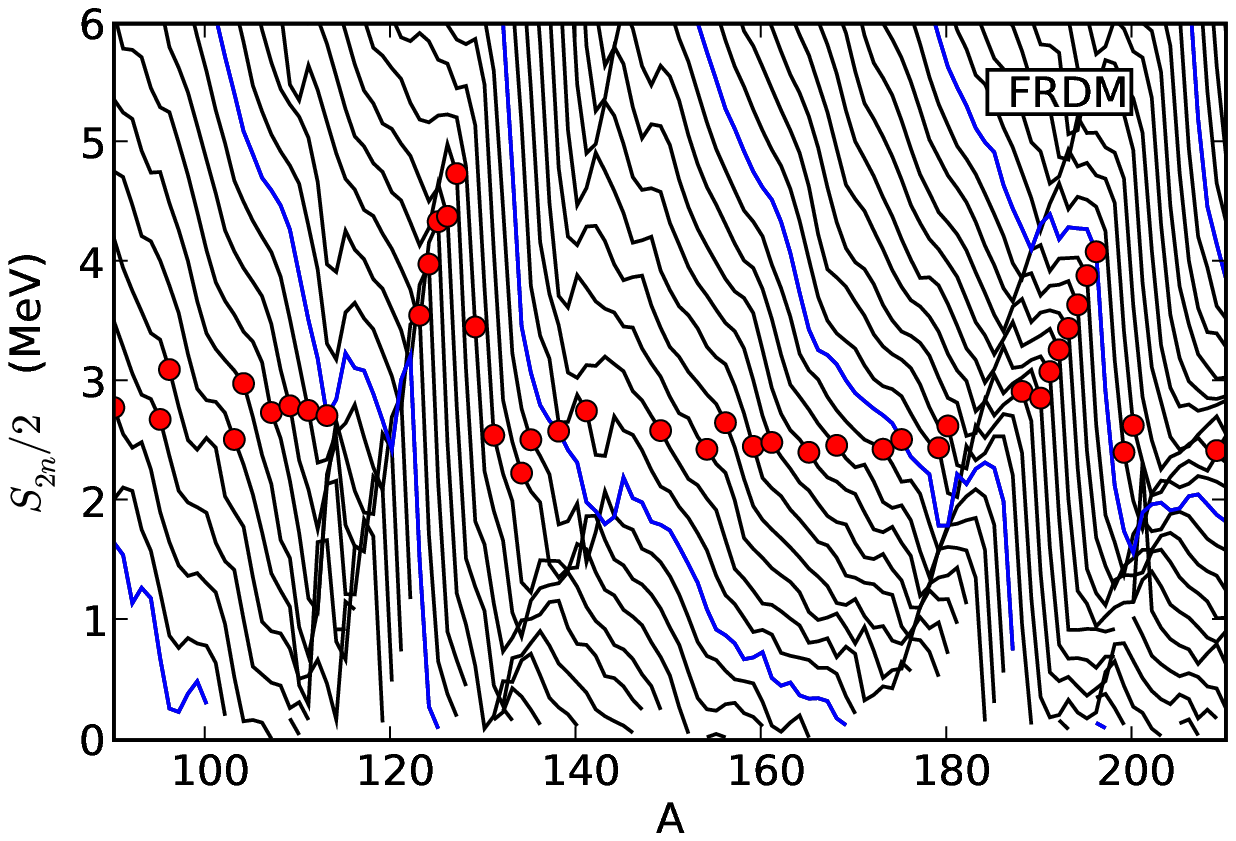}%
  \includegraphics[width=0.51\linewidth,angle=0]{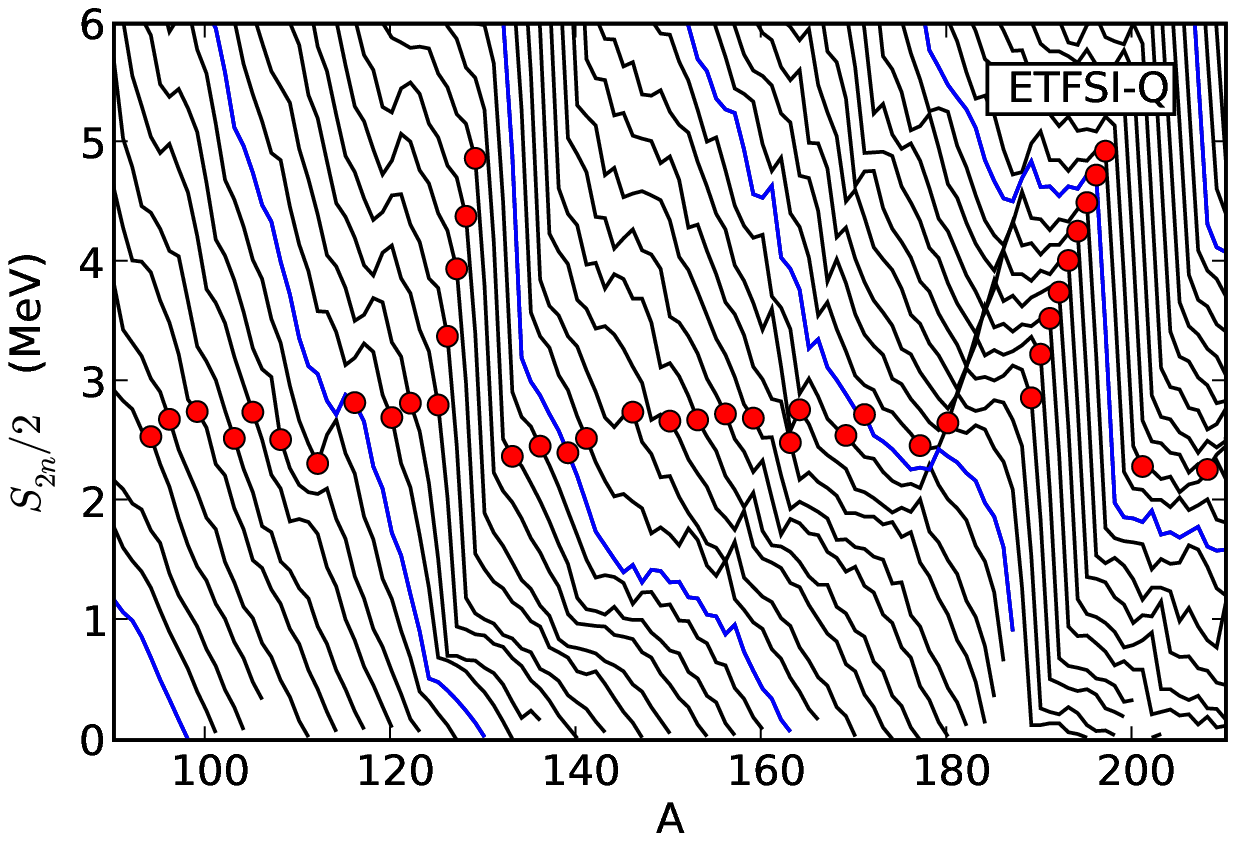}\\
  \includegraphics[width=0.51\linewidth,angle=0]{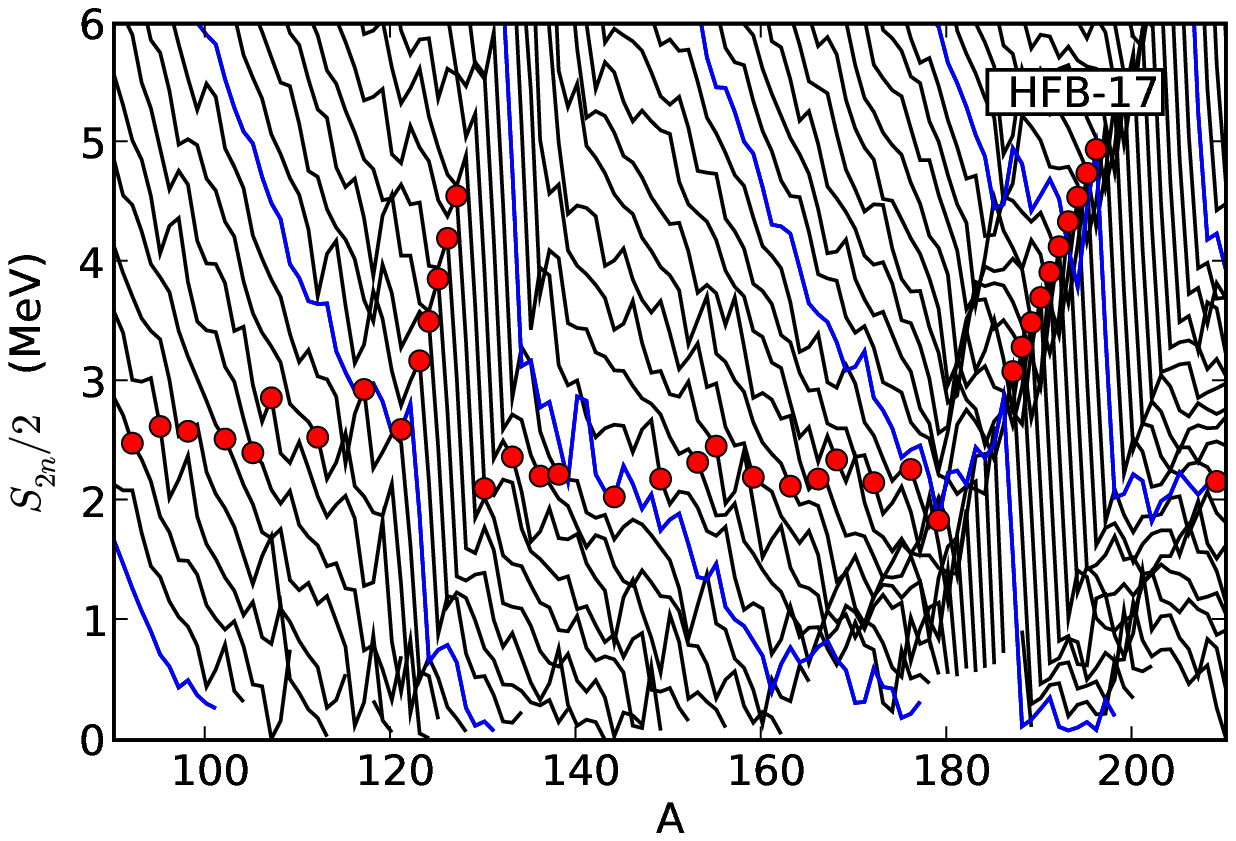}%
  \includegraphics[width=0.51\linewidth,angle=0]{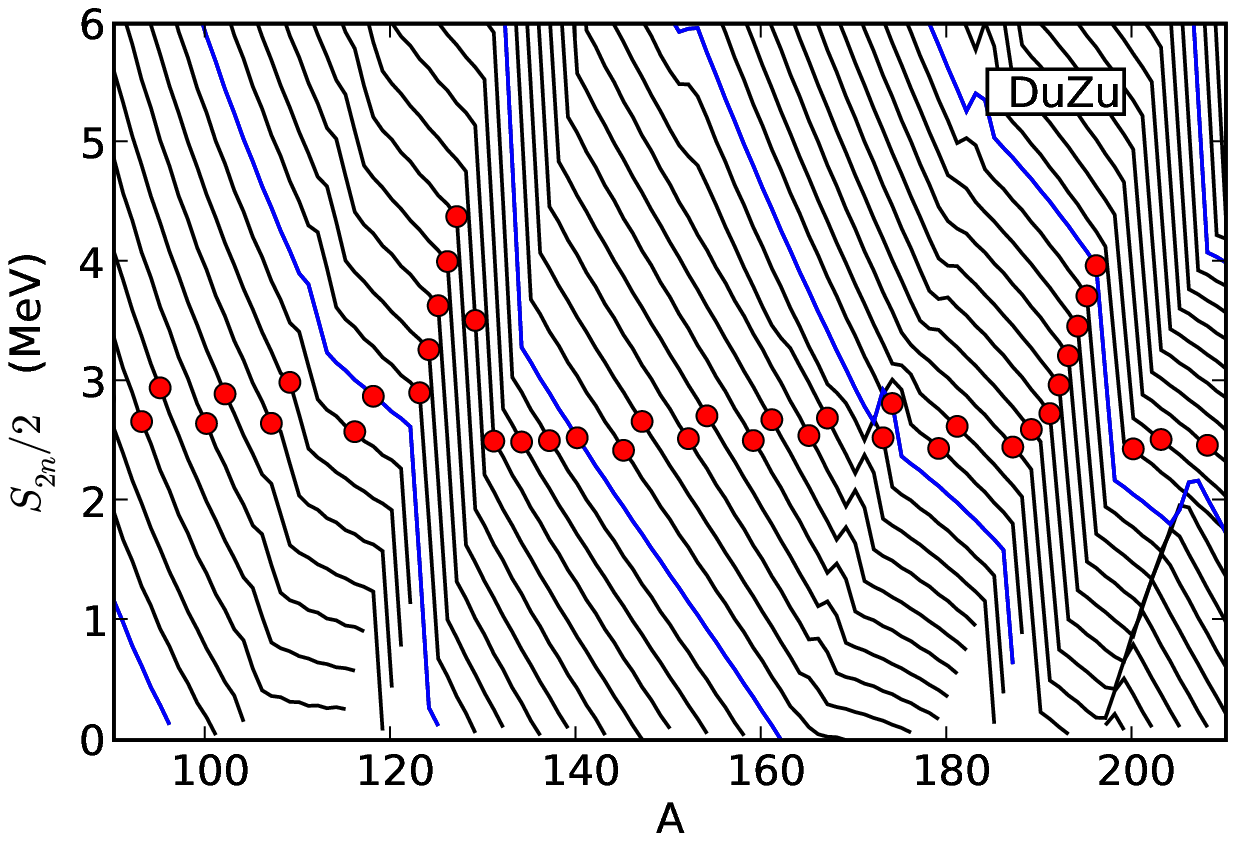}
  \caption{Two neutron separation energy contours for constant proton
    number increasing in steps of 1 vs. mass number. Blue lines
    represent Z~=~30,~40,~50,~60. The r-process path is shown at
    freeze-out ($Y_n/Y_{\mathrm{seed}}=1$) by dots.}
  \label{fig:s2npath}
\end{figure}

\begin{figure}[!htp]
  \includegraphics[width=0.5\linewidth,angle=0]{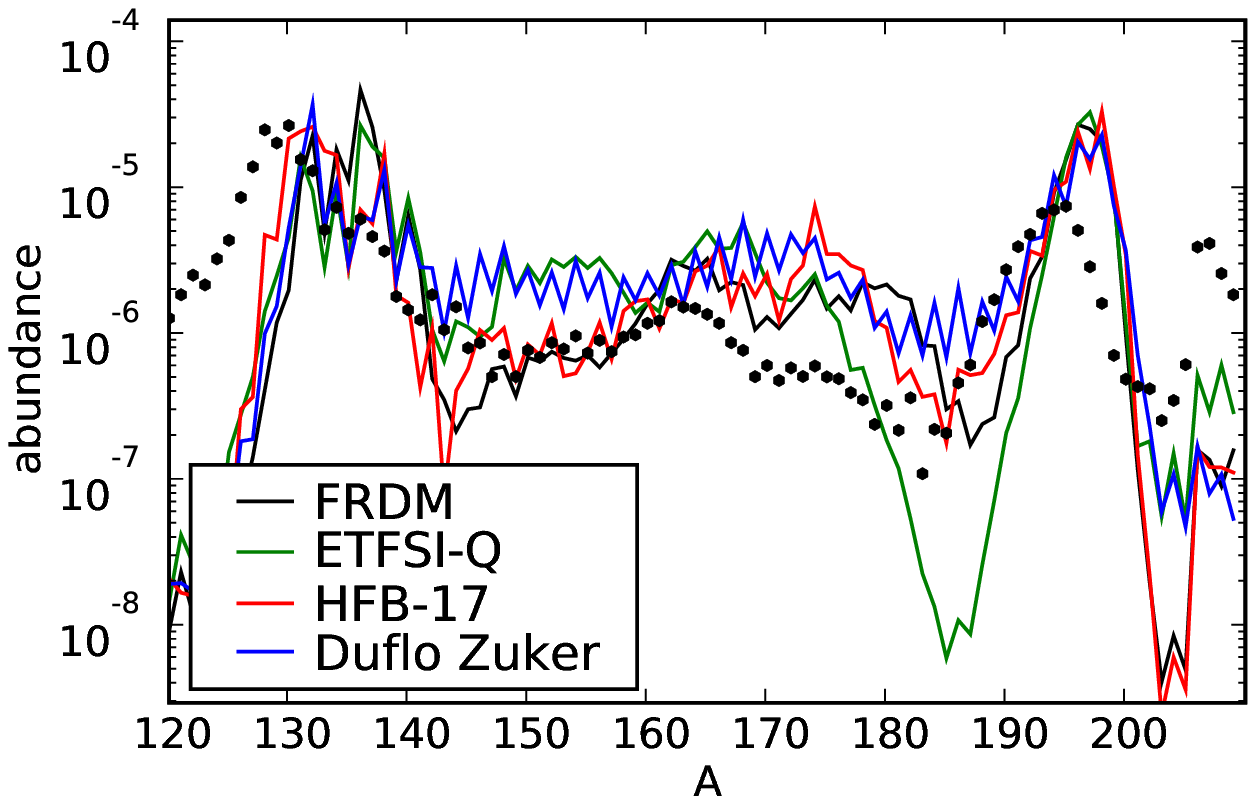}%
  \includegraphics[width=0.5\linewidth,angle=0]{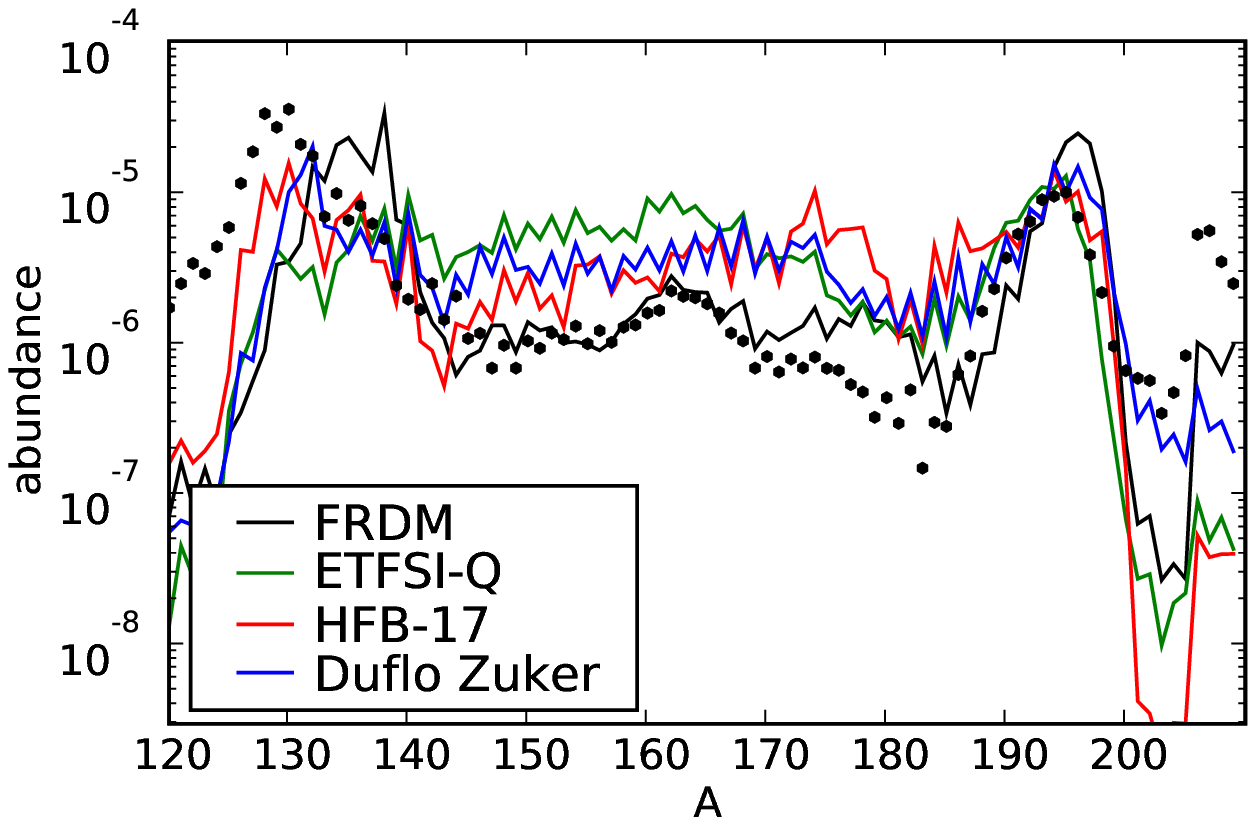}
  \caption{Abundances for the mass models indicated in the caption and
    for the equilibrium (left) and non-equilibrium (right) evolutions,
    compared to solar (dots).}
  \label{fig:abundmm}
\end{figure}

The sensitivity of the mass model have been investigated by
consistently changing neutron separation energies and neutron capture
rates for the mass models: FRDM~\cite{Moeller.Nix.ea:1995},
ETFSI-Q~\cite{Pearson.Nayak.Goriely:1996},
HFB-17~\cite{Goriely.Chamel.Pearson:2009}, and
Duflo-Zuker~\cite{Duflo.Zuker:1995}.  The presence and position of
peaks and trough in the abundances depends on features of the two
neutron separation energy ($S_{2n}$) shown in
Fig.~\ref{fig:s2npath}. When $S_{2n}$ abruptly drops for increasing
$N$, matter accumulates leading to the formation of peaks. While in
regions where $S_{2n}$ is flat or presents a saddle point behaviour,
several neutron captures occur almost instantaneously leaving a trough
in the abundances. This feature of the two neutron separation energy
is present in all mass models before $N=126$ as nuclei change from
deformed to spherical. In the equilibrium evolution, where
photodissociation is very important, this leads to the formation of a
big trough in the abundances at the moment neutrons are almost
exhausted, i.e. when $Y_n/Y_{\mathrm{seed}}=1$. Afterwards, as matter
decays to stability, neutron capture can fill up this trough or make
it bigger as it occurs in the abundances based on ETFSI-Q. Our results
shown in Fig.~\ref{fig:abundmm} present also some behaviours that are
characteristic of every mass model. In ETFSI-Q the quench before
$N=82$ leads to a slow down of the evolution and to a delayed
freeze-out. Moreover, the fluctuation of $S_{2n}$ before $N=126$ in
this mass model makes the trough around $A\approx 185$ bigger due to
neutron captures when matter moves back to stability. Results based on
FRDM are clearly affected by the anomalous behaviour of $S_{2n}$
before $N=90$, which produces the accumulation of matter and thus the
formation of peaks around $A\approx 135$ even in the non-equilibrium
evolution (Fig.~\ref{fig:abundmm}).

\subsection{Way back to stability}
\label{sec:back}
The abundances at freeze-out present a lot of fluctuations while the
final ones, after decay to stability, are smooth like the solar system
abundances. In the classical r-process calculation (waiting point
approximation) this is explained by beta-delayed neutron emission (see
e.g., \cite{Kratz.Bitouzet.ea:1993}). However, in dynamical r-process
calculations also neutron captures contribute to redistribution of
matter. The neutron captures become very important after freeze out,
when only few neutrons are available and nuclei compete to capture
them.  We find that the rare earth peak is due to neutron captures
when matter moves back to stability, as suggested in
Ref.~\cite{Surman.Engel.ea:1997}. This implies that the freeze out
cannot be very fast because neutrons are still needed to form this
feature which is present in the solar r-process abundances.

\begin{figure}[!htp]
  \includegraphics[width=0.5\linewidth,angle=0]{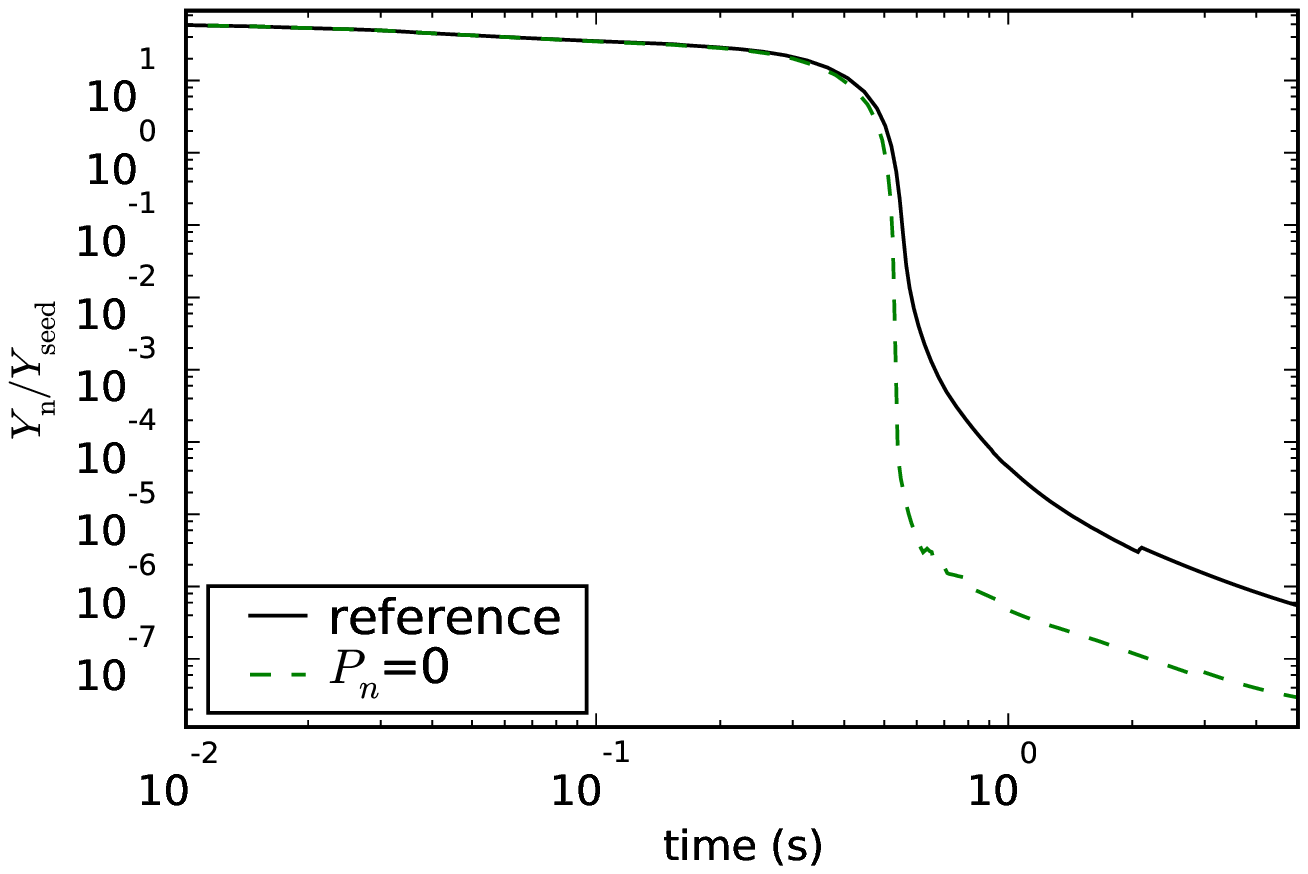}%
  \includegraphics[width=0.5\linewidth,angle=0]{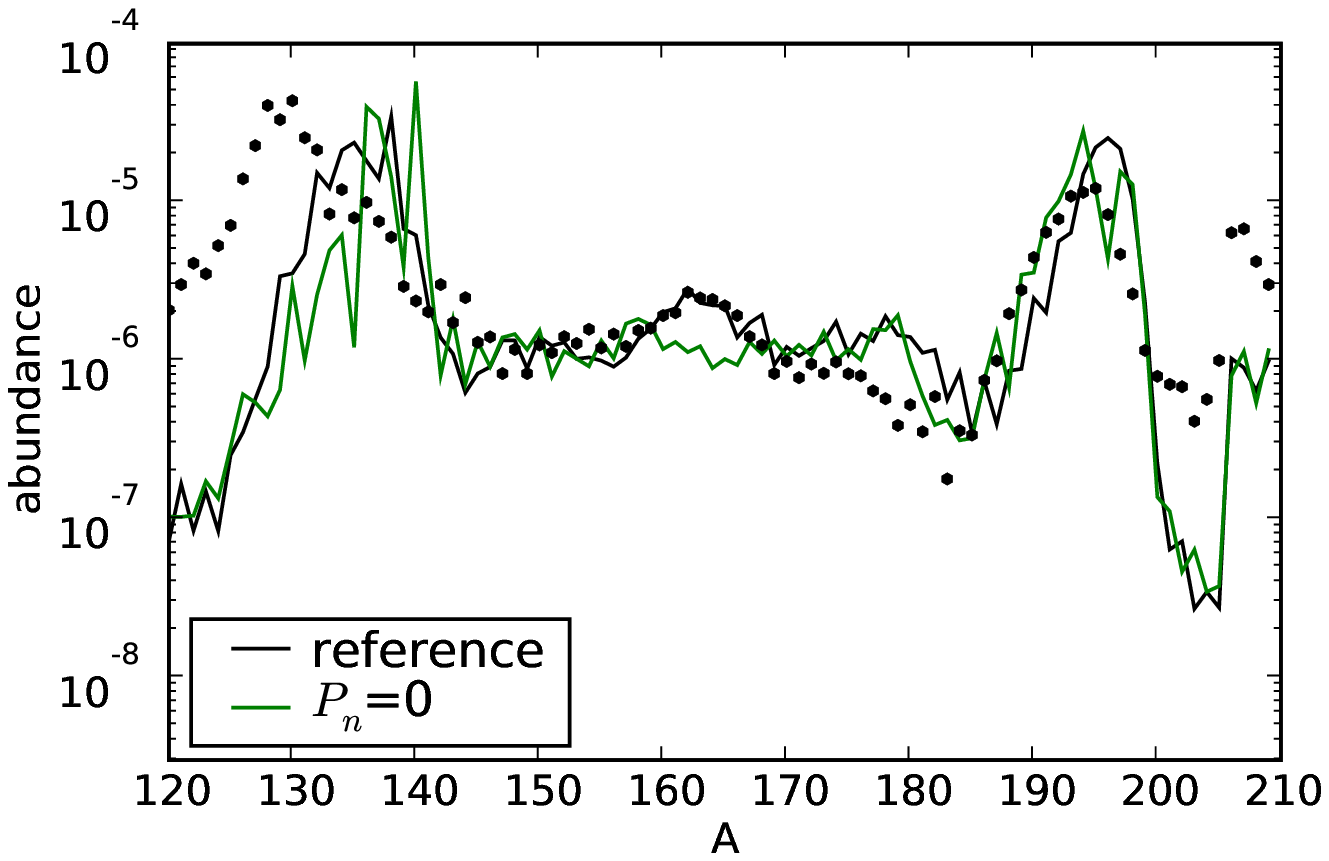}
  \caption{Neutron-to-seed ratio and abundances for the
    non-equilibrium evolution.  The black lines are for the reference
    case which is calculated with the standard nuclear input where
    neutrons are emitted with given probability ($P_n$) after beta
    decay. The green lines are for the case where $P_n=0$, therefore A
    is conserved during beta-decay.}
  \label{fig:betan}
\end{figure}

Finally, we found that the main contribution of the beta-delayed
neutron emission is the supply of neutrons. In our equilibrium
evolution there are almost no difference in the abundance calculated
with and without beta-delayed neutron emission. Since temperature are
high, photodissociation prevents the path to reach the regions far
from stability where the probability of emitting neutrons after beta
decay is higher. In contrast, in the non-equilibrium evolution the
neutron density is significantly smaller when no beta-delayed neutron
emission is assumed. The evolution of the neutron-to-seed ratio is
shown in the left panel of Fig.~\ref{fig:betan} for the calculations
with and without beta-delayed neutron emission. The significant
smaller neutron-to-seed ratio, when non beta-delayed neutron emission
is considered (green line), leads to less shift of the third peak
after freeze-out but also inhibits the formation of the rare earth
peak (right panel in Fig.~\ref{fig:betan}).

\section{Conclusions}
\label{sec:conclusions}
Recent long-time supernova simulations do not produce r-process
elements because the wind entropy is too low and the electron fraction
high, even staying proton rich during several seconds
\cite{Huedepohl.etal:2010}. However, the LEPP elements can be produced
as we have shown by comparing for the fist time the LEPP pattern in
UMP stars and integrated nucleosynthesis calculations based on
hydrodynamical wind simulations \cite{Arcones.Montes:2010}. In
proton-rich winds the LEPP pattern is very robust and reproduces
observed abundances from UMP stars. Neutron-rich winds are necessary
to explain the LEPP isotopes found in the solar system abundances, but
they do not lead to a robust pattern and overproduced nuclei around
A=90. This suggests that only a small fraction of the supernovae or of
the mass ejected by them can be neutron rich. Future observations of
isotopic abundances in ultra-metal poor stars could constrain the
evolution of the electron fraction in the neutrino-driven winds and
thus the neutrino properties (energy and luminosity).

The impact of the long-time dynamical evolution and of nuclear masses
on the r-process abundances can be still studied based on current
simulations by artificially increasing the entropy. This mimics the
hydrodynamical conditions of a neutrino-driven wind where the
r-process does occur. We have found that the relevance of the
different nuclear physics inputs depends on the long-time dynamical
evolution \cite{Arcones.Martinez-Pinedo:2010}. If an
$(n,\gamma)$-$(\gamma,n)$ equilibrium is reached, nuclear masses have
a big influence on the final abundances. While for a cold r-process
there is a competition between neutron capture and beta decay and
these two process become relevant.  This rises the importance of
future experiments to measure nuclear masses that will provide a
direct input for network calculations and constraints for the
theoretical mass models.

In both types of evolutions as matter decays to stability, our results
show that neutron captures are key to understand the final
abundances. Moreover, we found that beta-delayed neutron emission is
important not only for the redistribution of matter, but also for the
supply of neutrons. The late neutron captures are necessary to explain
features in the solar system abundances, such as the rare earth
peak. More experimental effort is necessary to test the validity of
the current theoretical cross sections and more sensitivity studies of
the impact of the neutron capture rates on the final abundances will
give rise to new insights.

\ack I would like to thank my collaborators, G.~Mart\'inez-Pinedo and
F.~Montes, for the contributions to the work presented here. Support
of Swiss National Science Foundation is acknowledged.

\section*{References}

\providecommand{\newblock}{}

\end{document}